\documentclass[12pt]{article}

\usepackage{cite}
\usepackage{epsfig}
\usepackage{graphics}
\usepackage{inputenc}
\inputencoding{latin1}

\def\kt{$K_\perp$}

\def\lum{$\mathcal{L}$}

\begin{document}

\sloppy

\begin{titlepage}

{\par\raggedleft
\texttt{February} \texttt{2007}\\
CERN-PH-TH/2007-019\par}
\bigskip{}

\bigskip{}
{\par\centering \textbf{\large Reconstructing Sparticle Mass Spectra using 
Hadronic Decays}
\large \par}
\bigskip{}

{\par\centering J. M. Butterworth$^{1}$, John Ellis$^{2}$ and A. R. 
Raklev$^{2,3}$\\
\par}
\bigskip{}

{\par\centering
{\small $^1$ Department of Physics \& Astronomy, University College 
London}\\
{\small $^2$ Theory Division, Physics department, CERN}\\
{\small $^3$ Department of Physics \& Technology, University of Bergen}\\
\par}

\bigskip{}

\begin{abstract}
\noindent
Most sparticle decay cascades envisaged at the Large Hadron Collider
(LHC) involve hadronic decays of intermediate particles. We use
state-of-the art techniques based on the \kt\ jet algorithm to
reconstruct the resulting hadronic final states for simulated LHC
events in a number of benchmark supersymmetric scenarios. In
particular, we show that a general method of selecting preferentially
boosted massive particles such as $W^\pm, Z^0$ or Higgs bosons
decaying to jets, using sub-jets found by the \kt\ algorithm,
suppresses QCD backgrounds and thereby enhances the observability of
signals that would otherwise be indistinct.  Consequently,
measurements of the supersymmetric mass spectrum at the per-cent level
can be obtained from cascades including the hadronic decays of such
massive intermediate bosons.
\end{abstract}


\end{titlepage}

\section{Introduction}
\label{sec:intro}

In a number of commonly considered supersymmetric (SUSY) models,
strongly-interacting sparticles will be abundantly pair-produced at
the Large Hadron Collider (LHC), and the resulting events containing
jets and missing energy will stand out above the Standard Model
background. The next challenge will be to make sense of the events,
which will typically involve cascade decays through several
intermediate sparticles. It will be particularly desirable to obtain
information on the SUSY masses and branching ratios. It has been
demonstrated that cascades involving lepton emission often provide
relatively clean signals, which can be used to reconstruct several
sparticle masses in some favourable benchmark scenarios, see
e.g.~\cite{Hinchliffe:1996iu,edges}. However, these decay channels
often suffer from branching ratios that are low compared to hadronic
decay modes, reducing the available sample size and therefore
restricting access to high-mass sparticles. Moreover, the charged
leptons are often associated with neutrinos, for example in chargino
($\chi^\pm$) decays. The presence of the neutrinos would confuse the
interpretation of the missing-energy signal provided by escaping
neutralinos in scenarios with $R$-parity conservation, as we assume
here, and make the reconstruction of chargino masses difficult. Some
progress on techniques for measuring chargino masses has been reported
for scenarios with one leptonic and one hadronic decay
chain~\cite{Nojiri:2003tv}, but these require models with a favourable
combination of the corresponding branching ratios. The ability to
reconstruct purely hadronic cascade decays would facilitate the
discovery and measurement of charginos in more general cases, as well
as enable heavier sparticle masses to be reconstructed.

In this paper, we take a new approach to the reconstruction of
fully-hadronic sparticle events, starting from the missing-energy
signal provided by the neutralino in $R$-conserving models, which
suppresses Standard Model (SM) backgrounds. We use jet analysis
techniques based on the \kt\ algorithm to identify the hadronic decays
of massive bosons such as the $W^\pm$, $Z^0$ or Higgs boson ($h$).  In
general, the majority of sparticle cascade decays yield fully-hadronic
final states containing such massive SM particles decaying to $q -
{\bar q}$ jet pairs~\cite{DeRoeck:2005bw}. A certain fraction of these
bosons, dependent on the sparticle masses in the scenario considered,
are highly boosted and the jets from the decays are closely aligned,
indeed overlapping, in pseudo-rapidity ($\eta$) - azimuthal angle
($\phi$) space. Such a situation presents both a challenge and an
opportunity. The challenge is that the choice of jet finder becomes
crucial: the treatment of jet overlaps can be sensitive to rather soft
and subtle QCD effects, and yet somehow one needs to retain the
information that the jet arises from the two-body decay of a massive
particle. The opportunity is that, since the jets are close together,
there is no combinatorial background; no need, for example, to combine
all pairs of jets to see if they reconstruct to the $W$ mass.

We address the reconstruction issue using the sub-jet technique that
was proposed previously as a way to identify high-energy $WW$
events~\cite{Butterworth:2002tt}. This technique and its extension to
hadronic sparticle decays are described in
Section~\ref{sec:subjet}. Then, in Section~\ref{sec:benchmarks} some
specific benchmark SUSY scenarios and the specific decay chains of
interest are described: these involve typical decays of intermediate
heavy charginos or neutralinos such as $\tilde\chi^\pm_1\to
W^\pm\tilde\chi^0_1$ and $\tilde\chi^0_2\to Z^0, h\tilde\chi^0_1$. The
simulations and analysis method are described in
Section~\ref{sec:analysis}, and the subsequent Section contains the
results and conclusions.

\section{Jet Finding and Sub-jets}
\label{sec:subjet}

The identification of jets which originate from a decaying massive
particle begins by using a jet algorithm to define the jet. In this
analysis the \kt\ algorithm~\cite{Catani:1993hr} is used, in the
inclusive mode with the $E$ reconstruction scheme. For each particle
$k$ and pair of particles $(k,l)$, the algorithm calculates the
quantities
\begin{eqnarray}
d_{kB} & = & p_{Tk}^2, \nonumber \\
d_{lB} & = & p_{Tl}^2, \nonumber \\
d_{kl} & = & \min(p^2_{Tk},p^2_{Tl})R^2_{kl}/R^2,
\end{eqnarray}
where $p_{Tk}$ is the transverse momentum of particle $k$ with respect
to the beam axis and
\begin{equation}
R^2_{kl}=(\eta_k-\eta_l)^2+(\phi_k-\phi_l)^2.
\end{equation}
The parameter $R$ is a number of order one. If it is set below unity,
it is less likely that a given particle will be merged with a jet,
which in turn leads to narrower jets. Thus $R$ plays a somewhat
similar role to the adjustable cone radius in cone algorithms. In this
analysis, initial studies indicated that $R = 0.7$ provided a good
compromise between efficiency and the mass resolution, and this value
is used throughout. Returning to the algorithm: all the $d$ values are
then ordered. If $d_{kB}$ or $d_{lB}$ is the smallest, then particle
$k$ or $l$ is labelled a jet and removed from the list. If $d_{kl}$ is
the smallest, particles $k$ and $l$ are merged by adding their
four-momenta. The list is recalculated and the process is repeated
until the list is empty. Thus the algorithm clusters all particles
into jets, and a cut on transverse momentum can then be used to select
the hardest jets in the event. The algorithm is infrared safe, and has
the additional benefit that each particle is uniquely assigned to a
single jet. Recently a fast implementation of the algorithm has been
developed~\cite{fastjet} which makes it practical for use even in the
very high multiplicity events expected at the LHC.

In selecting a candidate for a hadronic decay of the $W^\pm, Z^0$ or
$h$, first cuts on the $p_T$ and the pseudo-rapidity ($\eta$) of the
jet are applied so as to ensure that they are energetic enough to
contain a boosted heavy-particle decay and are in a region of good
detector acceptance. A cut is then applied on the mass of the jet
(calculated from the four-vectors of the constituents) to ensure that
it is in a window around the nominal mass of the desired particle.

The next step is to force the jet to decompose into two sub-jets. The
main extra piece of information gained from the sub-jet decomposition
is the $y$ cut at which the sub-jets are defined: $y \equiv
d_{kl}/(p_T^{\rm jet})^2$, where $p_T^{\rm jet}$ is the transverse
momentum of the candidate jet containing the sub-jets $k$ and $l$. In
the case of a genuine $W^\pm, Z^0$ or $h$ decay, the expectation is
that the scale at which the jet is resolved into sub-jets
(i.e., $yp_{T}^{2}$) will be ${\cal {O}}(M^{2})$, where $M$ is the
$W^\pm, Z^0$ or $h$ mass. For QCD jets initiated by a single quark or
gluon, the scale of the splitting is expected to be substantially
below $p_T^2$, i.e., $y \ll 1$, since in the region around the jet
strongly-ordered DGLAP-like~\cite{DGLAP} QCD evolution dominates.

\begin{figure}
\begin{center}
\epsfxsize 15cm
\epsfbox{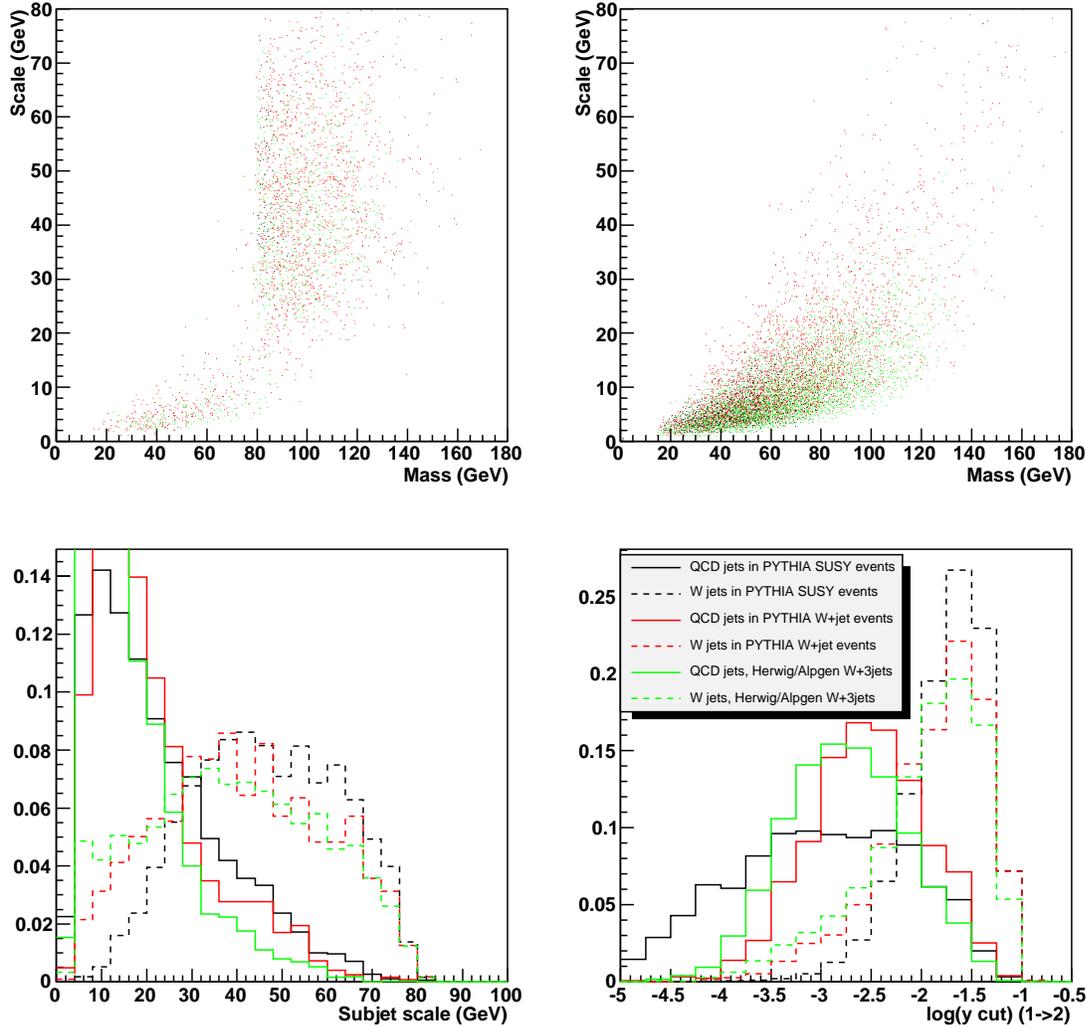}
\caption{Scatter plot of the jet mass against the jet splitting scale
$yp_T^2$ for (a) jets from $W^\pm$ decays (determined by a match better than
0.1 units in $\eta-\phi$) and (b) QCD jets in $W^\pm$+jet and SUSY
events. The distributions of the splitting scale are shown in (c) and
the $y$ distributions in (d), for the same types of jets, after a cut
on the jet mass at $75 < M < 90$~GeV. In all these plots, the
requirement $p_T > 250$~GeV is applied to all jets. The histograms in
the lower two plots are normalised to unity.}
\label{fig:scale}
\end{center}
\end{figure}

This distinction provides new information in addition to the jet mass
itself, as is illustrated in Figures~\ref{fig:scale}a
and~\ref{fig:scale}b, where the correlation between the jet mass and
the splitting scale is shown for $W^\pm$ jets and QCD jets
respectively. The events shown are $W^\pm$+jet events and SUSY events
generated using {\sc Pythia 6.408}~\cite{pythia}, and $W^\pm$+3jet
events generated using ALPGEN~\cite{alpgen} for the matrix element,
HERWIG~6.510~\cite{herwig} for parton showering and {\sc
Jimmy}~\cite{jimmy} for the underlying event. The parameters for the
underlying event and the parton showers were those of the ATLAS tune
of {\sc Pythia} and the CDF tune A of {\sc HERWIG} and {\sc Jimmy},
taken from~\cite{tunes}~\footnote{More details on the SUSY event
generation are given in Section~\ref{sec:analysis}.}. These models
have been shown to give a good description of a wide variety of
data. In particular, the modelling of the internal jet structure by
leading-logarithmic parton showers is known to be good for jets
produced in $p\bar{p}$ collisions~\cite{tevjets}, $ep$ collisions and
photo-production~\cite{herajets}, and in $e^+e^-$ annihilation events
and $\gamma\gamma$ collisions~\cite{lepjets}.

Although there are some differences between the results of the {\sc
Pythia} and HERWIG/ALPGEN simulations, the conclusions are
similar. Fig.~\ref{fig:scale}a confirms that the splitting scale in
$W^\pm$ decays is relatively large, typically $\ge 20$~GeV, whereas
Fig.~\ref{fig:scale}b shows that the splitting scale for QCD jets with
masses $\sim 80$~GeV is typically $\le 20$~GeV.  The distributions of
of $p_{T}\sqrt{y}$ and of $y$ are shown in Figures~\ref{fig:scale}c
and~\ref{fig:scale}d for $W^\pm$ jets and QCD jets. The distributions
are qualitatively similar, whether they are generated in SUSY events
or in conventional $W^\pm$+jets events, and whether (in the latter
case) they are generated using {\sc Pythia} or HERWIG. The scale of
the splitting is seen to be peaked close to the $W^\pm$ mass for
genuine $W^\pm$ decays, whichever the environment in which they are
produced, and to be softer for QCD jets which just happen to
reconstruct to the $W^\pm$ mass, again in both environments.

This distinction was noted and successfully used to identify $W^\pm$
decays in simulations of $W^+W^-$ scattering at LHC energies
in~\cite{Butterworth:2002tt}. Here, this information is used in the
analysis described in Section~\ref{sec:analysis} to refine the
selection of $W^\pm, Z^0$ and $h$ candidates in sparticle decay
cascades. Although no detector simulation is employed in this
analysis, the cut on the scale has been shown previously to be robust
against the effects of the underlying
events~\cite{Butterworth:2002tt}, as well as effects due to the
calorimeter granularity and resolution~\cite{theses}.

\section{SUSY Decay Chains and Benchmark Scenarios}
\label{sec:benchmarks}

The pair-production of heavy sparticles in generic SUSY models yields
on-shell electroweak gauge bosons or Higgs bosons via the decays of
heavy gauginos that appear as intermediate steps in cascades, e.g.,
${\tilde q}_L \to \tilde\chi^0_2 q$ or ${\tilde
q}_L\to\tilde\chi^\pm_1 q'$, followed by $\tilde\chi^0_2 \to Z^0, h +
\tilde\chi^0_1$ or $\tilde\chi^\pm \to W^\pm +
\tilde\chi^0_1$ decays.  Exceptions are cases with gaugino masses that are
nearly degenerate, in which there may be large branching ratios for
decays via off-shell heavy bosons. Since the largest branching ratios
for $W^\pm, Z^0$ and $h$ decays are those into hadronic ${\bar q} q$
final states, purely hadronic final states dominate in cascade decays
via on-shell bosons, and these are also potentially important in
off-shell decays.

\begin{figure}
\begin{center}
\psfig{file=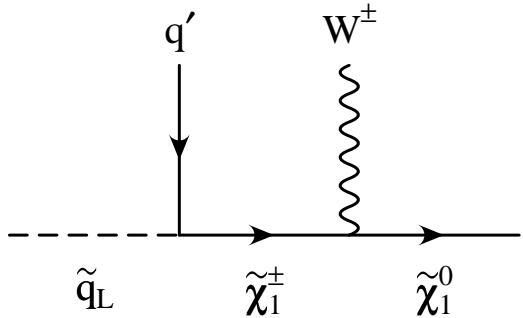,width=7.0cm}
\caption{The SUSY decay chain of Eq.~(\ref{eq:chargino}).}
\label{fig:qW} 
\end{center}
\end{figure}

For the decay chain
\begin{equation}
\tilde q_L \rightarrow \tilde\chi_1^\pm q' \rightarrow \tilde\chi_1^0 
W^\pm q'
\label{eq:chargino}
\end{equation}
shown in Fig.~\ref{fig:qW}, one can demonstrate that the invariant
mass distribution of the quark-$W$ system has a minimum and maximum
given by
\begin{equation}
(m_{qW}^{\rm{max}\slash\rm{min}})^2=
m_W^2+\frac{(m_{\tilde q_L}^2-m_{\tilde\chi_1^\pm}^2)}{m_{\tilde\chi_1^\pm}}
(E_W\pm|\vec{p}_W|),
\label{eq:endqW}
\end{equation}
where
\begin{equation}
|\vec{p}_W|^2=
\frac{(m_{\tilde\chi_1^\pm}^2-m_{\tilde\chi_1^0}^2-m_W^2)^2-4m_{\tilde\chi_1^0}^2m_W^2}{4m_{\tilde\chi_1^\pm}^2}
\label{eq:pW}
\end{equation}
is the $W$ momentum in the chargino rest frame. If measurable, these
endpoints give a model-independent relationship between the three SUSY
masses, modulo the existence of the decay chain. If both endpoints can
be determined experimentally, the squark mass can be eliminated,
giving the chargino mass in terms of the lightest
neutralino. Supplementary model-dependent assumptions, such as the
relationship $m_{\tilde\chi_1^\pm} \approx 2m_{\tilde\chi_1^0}$ that
holds approximately in the constrained MSSM (CMSSM), can then be used
to determine individual masses of all three sparticles in specific
theoretical frameworks.

The decay chains
\begin{equation}
\tilde q_L \rightarrow \tilde\chi_2^0 q \rightarrow \tilde\chi_1^0 h q
\label{eq:chi02h}
\end{equation}
and
\begin{equation}
\tilde q_L \rightarrow \tilde\chi_2^0 q \rightarrow \tilde\chi_1^0 Z^0 q,
\label{eq:chi02Z}
\end{equation}
involving the second lightest neutralino, have endpoint formulae with
the same structure, given by the substitutions
\begin{equation}
W\rightarrow h, \quad \tilde\chi_1^\pm\rightarrow\tilde\chi_2^0
\end{equation}
and
\begin{equation}
W\rightarrow Z, \quad \tilde\chi_1^\pm\rightarrow\tilde\chi_2^0,
\end{equation}
respectively, in Eqs.~(\ref{eq:endqW}) and (\ref{eq:pW}).

In~\cite{DeRoeck:2005bw} benchmark models with such cascade decays
were explored in a more generic setting than that allowed in the
CMSSM. Relaxing the GUT-scale universality of the Higgs scalar masses
or introducing a gravitino dark matter candidate admits values of the
(supposedly universal) scalar squark and slepton mass $m_0$ outside
the narrow range allowed by the cold dark matter
density~\cite{Bennett:2003bz,Spergel:2003cb} within the restrictive
CMSSM framework. This, in turn, allows for a wide range of values for
the branching ratios of the lightest chargino and the next-to-lightest
neutralino into $W^\pm, Z^0$ and $h$.  In our study, we use the
benchmark points $\alpha$, $\beta$, $\gamma$ and $\delta$
from~\cite{DeRoeck:2005bw} to illustrate these possibilities. The
points $\alpha$, $\beta$, $\gamma$ are scenarios with relatively light
SUSY masses, that should be easy to probe at the LHC, while $\delta$
represents a more challenging scenario with lower cross sections. Common to
all four points is the large branching ratio for squark to
chargino/neutralino decays, $\rm{BR}(\tilde q_L\to\tilde\chi_1^\pm
q)\approx 60$\% and $\rm{BR}(\tilde q_L\to\tilde\chi_2^0 q)\approx
30$\%. The corresponding chargino and neutralino bosonic branching
ratios are shown in Table~\ref{tab:BRs}.

\begin{table}\begin{center}
\begin{tabular}{cccc} \hline
Point/BR & $\tilde\chi_2^0\rightarrow\tilde\chi_1^0 Z$
& $\tilde\chi_2^0\rightarrow\tilde\chi_1^0 h$
& $\tilde\chi_1^\pm\rightarrow\tilde\chi_1^0 W^\pm$ \\ \hline
$\alpha$ & 98.6 &  0.0 & 99.6 \\
$\beta$  &  7.5 & 64.5 & 79.0\\
$\gamma$ &  0.0 &  0.0 & 99.9 \\
$\delta$ &  5.4 & 92.0 & 97.5 \\ \hline
\end{tabular}
\caption{Branching ratios for $\tilde\chi_2^0$ and $\tilde\chi_1^\pm$ in the
selected SUSY benchmark models. The decays are calculated using {\sc
SDECAY~1.1a}~\cite{Muhlleitner:2003vg}.}
\label{tab:BRs}
\end{center}
\end{table}

\begin{figure}
\begin{center}
\epsfxsize 15cm
\epsfbox{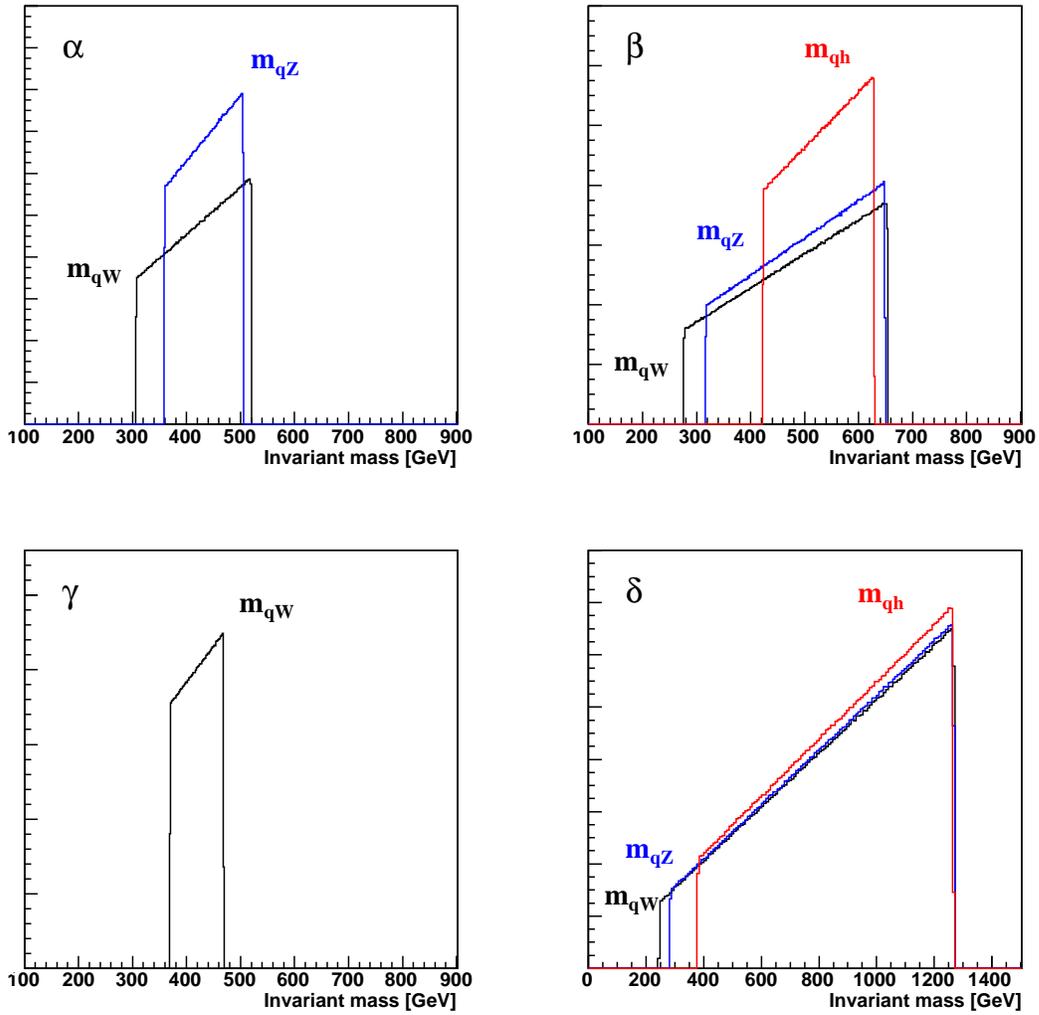}
\caption{The invariant mass distributions of $qW^\pm$, $qZ$ and $qh$
combinations for the chosen SUSY benchmark points.}
\label{fig:imtruth}
\end{center}
\end{figure}

We show the predicted shape of the invariant mass distributions for
the kinematically-allowed quark-boson combinations in the decay chains
(\ref{eq:chargino}), (\ref{eq:chi02h}) and (\ref{eq:chi02Z}), for all
the considered benchmark points, in Fig.~\ref{fig:imtruth}. These
distributions consider only the kinematics of the decay chain and
assume zero width for all particles. The distributions have a
characteristic trapezoidal shape with only small variations in the
location of the upper edge with the changing boson mass. This
similarity is due to the neutralino/chargino mass degeneracy typical
in CMSSM models, and demonstrates that the upper endpoint given by
Eq.~(\ref{eq:endqW}) is to a large extent insensitive to the boson
mass. Locating the upper endpoints for two different decay chains
involving the chargino and neutralino, respectively, would be an
excellent test of this mass degeneracy. If they are similar, this would point towards SUSY
scenarios where both the second-lightest neutralino and the lightest
chargino are `wino-like'.

\section{Simulation and Analysis}
\label{sec:analysis}

\subsection{Simulation}

In order to simulate sparticle pair-production events at the LHC, we
use {\sc Pythia~6.408}~\cite{pythia} with {\sc CTEQ 5L}
PDFs~\cite{Lai:1999wy} interfaced to {\sc HZTool}~\cite{hztool}, with
some minor changes to allow for simulations of SUSY scenarios. Decay
widths and branching ratios for the SUSY particles are calculated with
{\sc SDECAY~1.1a}~\cite{Muhlleitner:2003vg}. For the $\alpha$, $\beta$
and $\gamma$ benchmark points we simulate a number of SUSY events
equivalent to $30$~$\rm{fb}^{-1}$, giving results that should be
representative of the early reach of the LHC experiments at low
luminosity. For the $\delta$ benchmark point, with its considerably
higher masses and lower cross sections, we simulate a number of events
equivalent to $300$~$\rm{fb}^{-1}$, yielding results that should
indicate the ultimate reach of the LHC experiments with their design
luminosity.

We have also generated SM backgrounds with {\sc Pythia} in five $p_T$
bins from $p_T=50$~GeV to $7$~TeV. These samples rely on the parton
shower to simulate extra jets. This should be a reasonable
approximation in the important kinematic regions for some processes, such as $t \bar{t}$,
where the scale of the hard interaction is $>350$~GeV and we rely on
parton showers to simulate jets at around 150-200~GeV. However, this
is not adequate in all cases, and so in addition we have investigated
other possibly important sources of background, by using ALPGEN/HERWIG
to generate final states containing dibosons plus one or two jets, or
a single boson plus two or three jets~\footnote{Appropriate cuts,
considering the final-state selections to be used in
Section~\ref{sec:cuts}, have been applied to the partons to reduce the
amount of event generation required, and the ALPGEN parton-shower
matching scheme was used where appropriate.}. In these cases the jet
multiplicities are not well modelled by parton showers, but the
internal jet structure should still be well described. These
backgrounds turn out not to be very important for the $\alpha, \beta$
and $\gamma$ scenarios, but are found to be significant in the
$\delta$ scenario due to the small SUSY production cross section. For
the multiple jet processes there is a small amount of double counting
with the {\sc Pythia} samples, which we neglect here, giving a
conservative estimate of the backgrounds in this sense. The
generated numbers of events for the various processes, per $p_T$ bin
where used, are shown in Table~\ref{tab:efficiency} along with the
corresponding integrated luminosities.

\begin{table}\begin{center}
\begin{tabular}{lrrrr} \hline
Sample    & $N_{\rm{generated}}$ & \lum\ [fb$^{-1}$] & $N_{\rm{pass}}(\alpha-\gamma)$    & $N_{\rm{pass}}(\delta)$ \\ \hline
$t\bar t$ &            &       &  256.7 & 1287.0 \\
\ 50-150  & 26,500,000 &  93.0 &        &        \\
\ 150-250 & 10,000,000 &  95.6 &        &        \\
\ 250-400 &  3,500,000 & 120.0 &        &        \\
\ 400-600 &    500,000 & 129.6 &        &        \\
\ 600-    &    500,000 & 902.4 &        &        \\
$Wj$      &            &       &    5.2 &   34.5 \\
\ 50-150  &  1,100,000 &   0.1 &        &        \\
\ 150-250 &  1,100,000 &   2.9 &        &        \\
\ 250-400 &  1,100,000 &  20.2 &        &        \\
\ 400-600 &  1,100,000 & 154.3 &        &        \\
\ 600-    &    600,000 & 507.2 &        &        \\
$Zj$      &            &       &    3.2 &    3.0 \\
\ 50-150  &    100,000 &   0.0 &        &        \\
\ 150-250 &    100,000 &   0.6 &        &        \\
\ 250-400 &    100,000 &   4.3 &        &        \\
\ 400-600 &    100,000 &  32.7 &        &        \\
\ 600-    &    100,000 & 199.7 &        &        \\
$Wjj$     &    157,800 & 114.5 &   49.2 &  450.5 \\
$Zjj$     &    112,000 &  99.9 &   43.9 &  417.7 \\
$Wjjj$    &     50,300 & 227.9 &  127.8 & 1109.4 \\
$Zjjj$    &     27,300 & 156.6 &  194.4 & 1782.9 \\
$WW/WZ/ZZ$&            &       &    9.6 &   95.3 \\
\ 50-150  &    100,000 &   1.8 &        &        \\
\ 150-250 &    100,000 &  29.2 &        &        \\
\ 250-400 &    100,000 & 158.2 &        &        \\
\ 400-600 &    100,000 & 945.2 &        &        \\
\ 600-    &     10,000 & 437.0 &        &        \\
$WWj$     &    201,200 & 100.7 &    9.8 &   98.3 \\
$WZj$     &    162,400 &  90.2 &    0.0 &    0.0 \\
$ZZj$     &     69,500 & 426.5 &    2.3 &   17.6 \\ 
$WWjj$    &    107,300 &  98.7 &   23.4 &  215.8 \\ 
$WZjj$    &    179,000 & 248.4 &   55.2 &  455.5 \\ 
$ZZjj$    &     18,900 & 167.0 &    5.9 &   59.3 \\ \hline
\end{tabular}
\caption{The numbers of generated events, separated by $p_T$ bin
where used, the corresponding integrated luminosities
and the numbers of events passing the cuts for the $qW$ distribution, as
described in Section~\ref{sec:cuts}. Not shown are $2\to 2$ QCD
events, which are found not to contribute.}
\label{tab:efficiency}
\end{center}
\end{table}

No attempt to simulate detector effects has been made, but we have
included semi-realistic geometrical requirements for jets and leptons,
restricting ourselves to leptons with $|\eta|<2.5$ and jets with
$|\eta|<4.0$~\footnote{Details of the jet reconstruction were given in
Section~\ref{sec:subjet}.}.  We calculate the missing transverse energy
from the vector sum of the momenta of all the visible particles within
$|\eta|<5.0$, excluding neutrinos and neutralinos. For our
investigation of Higgs decays into $b\bar b$ pairs we use a simple
statistical model for identifying two collimated $b$-jets, by tagging
any jet matched to the direction of two $b$ quarks with $p_T > 15$~GeV
at the parton level~\footnote{For a match we require a distance to the jet of
less than 0.4 units in the $(\eta, \phi)$ plane for each $b$ quark.} as
resulting from two collimated $b$-jets with an efficiency of $40$\%,
and using a mis-tagging rate of $1$\% for all other jets. The numbers
used are a naive estimate from single $b$-jet tagging at the LHC. More
exact predictions for the performance of the LHC detectors on such
jets will require the full simulation of a detector, which is beyond
the scope of this study.

\subsection{Signal Isolation}
\label{sec:cuts}

In order to isolate events with the decay chain (\ref{eq:chargino}),
we use the following cuts:
\begin{itemize}
	\item Require missing energy $\not\!\!E_T>300$~GeV;
  	\item Require at least one $W^\pm$ candidate jet with
  	\begin{itemize}
		\item transverse momentum $p_T>200$~GeV,
		\item jet mass around the $W$ mass: $75<m_W<105$~GeV,
		\item separation scale $1.5<\log{(p_T\sqrt{y})}<1.9$;
	\end{itemize}
	\item Veto events with a top candidate, i.e.\,a jet-$W$ combination with invariant 
	mass in the range $150-220$~GeV.
	\item Require two additional jets with $p_T>200,150$~GeV;
	\item Veto events containing leptons ($e$ or $\mu$) with 
	      $p_T>10$~GeV.
\end{itemize}
The asymmetric cut on jet mass is due to the tendency of the jet
algorithm to overestimate the jet mass and energy by including
contributions from the underlying event and parton shower. Jets which
pass the jet mass cut for $W$ candidates are re-calibrated to the
known $W$ mass by rescaling the four-vector. To find the invariant
$qW^\pm$ mass in events that pass all the cuts we combine the $W^\pm$
candidate with any jet that passes the $p_T>200$~GeV requirement. This
creates some combinatorial background from signal events where we have
picked the wrong jet.

The remaining non-SUSY background is mainly a mix between $t\bar t$
events and vector boson (single and pair) production in association
with multiple jets. We show the surviving number of events for the
given cuts and integrated luminosities in
Table~\ref{tab:efficiency}. While this background is relatively
unimportant for the benchmarks with larger cross sections, it is
highly significant for the $\delta$ benchmark. Thus we have imposed
one further cut for this benchmark:
\begin{itemize}
	\item Require that the angle in the transverse plane between the
	      missing momentum and the $W$ candidate is larger than $\pi/8$.
\end{itemize}
The reason for this cut is that a significant fraction of the
surviving $t\bar t$ events feature the $W\to\tau\nu_\tau$ decay of a
highly boosted $W$, where the $\tau$ subsequently decays
hadronically. This gives large amounts of missing energy from
neutrinos, and the possibility of misidentifying the $\tau$ jet, or a
collimation of the $\tau$ jet and the $b$ jet from the same top as
the $W$ candidate. The result is a strong correlation between the
missing momentum and the direction of the $W$ candidate for the $t\bar
t$ background.

For the other two decay chains, (\ref{eq:chi02h}) and
(\ref{eq:chi02Z}), we use the same signal extraction procedure,
replacing the cut values for the jet mass cut and the separation scale
cut with appropriate values. For $Z$ candidates we require
$90<m_Z<115$~GeV and $1.6<\log{(p_T\sqrt{y})}<2.0$, while for the
Higgs boson we require $110<m_h<140$~GeV and
$1.8<\log{(p_T\sqrt{y})}<2.1$. In the Higgs case, we further require
$b$-tagging as described in the previous Section, while we do not
implement the lepton veto. For our mass reconstruction in
Section~\ref{sec:mass} we also look at the relatively clean signal of
leptonic $Z$ decays. We keep the cuts on missing energy and additional
jets and require two opposite-sign, same-flavour leptons with
$p_T>10$~GeV and an invariant mass in the range $85-95$~GeV.

\subsection{Invariant Mass Distributions}

\begin{figure}
\begin{center}
\epsfxsize 13.5cm
\epsfbox{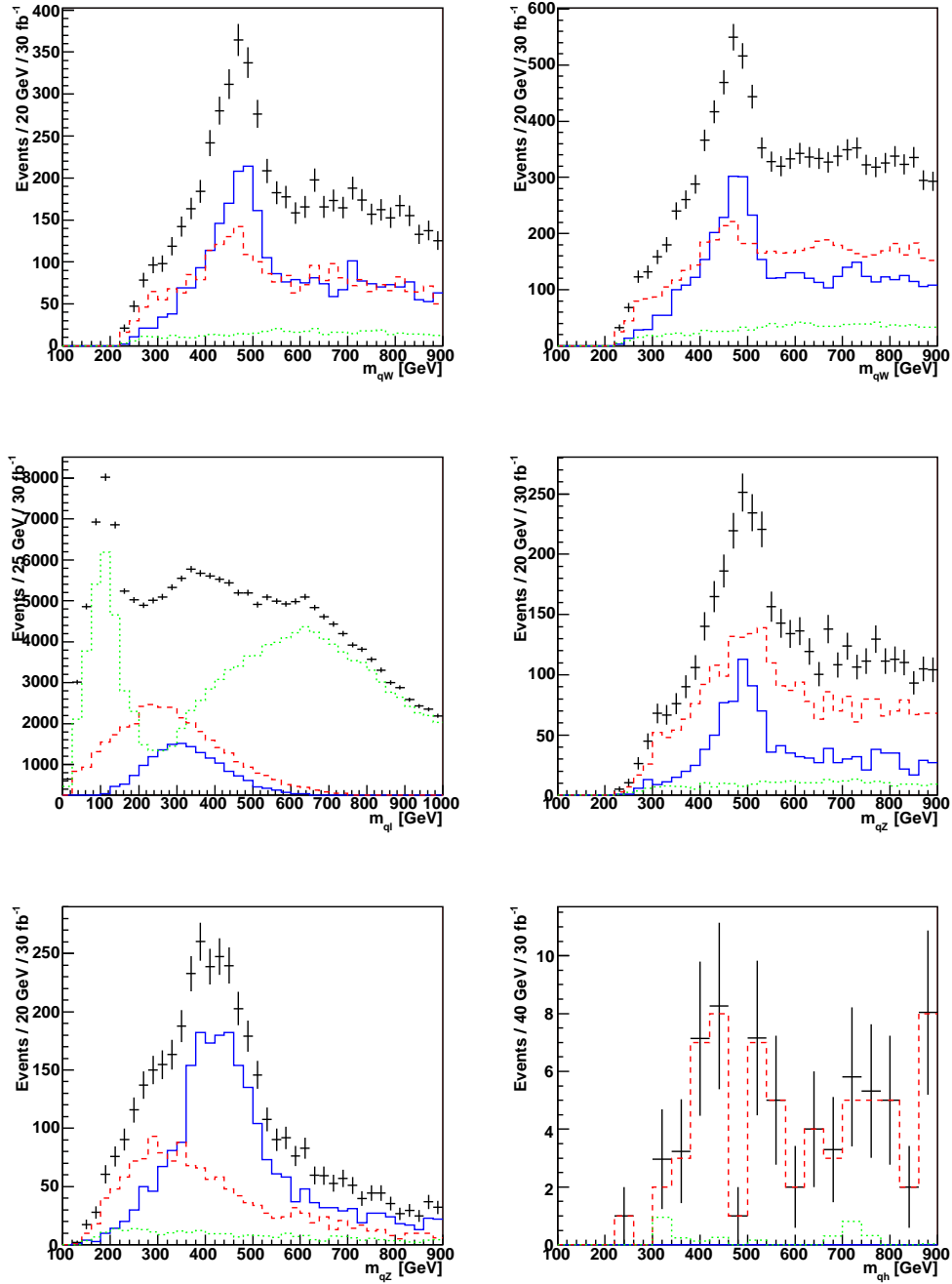}
\caption{Invariant mass distributions for SUSY benchmark scenario
$\alpha$: combinations of jets and $W^\pm$ candidates (a) with and (b)
without the cut on the separation scale, (c) the $\ell q$ invariant
mass distribution resulting from $W^\pm \to \ell^\pm \nu$ decays,
combinations of jets and $Z^0$ bosons decaying (d) hadronically and
(e) leptonically, and (f) combinations of jets and $h$ bosons. Signal
- blue, solid lines; SUSY background - red, dashed lines; SM
background - green, dotted lines.}
\label{fig:m_alpha}
\end{center}
\end{figure}

The resulting invariant $qW^\pm$ mass distribution for the $\alpha$
benchmark point is shown in Fig.~\ref{fig:m_alpha}a, for our
simulation of 30~fb$^{-1}$. There is a clear signal peak from events
that contain the decay chain (\ref{eq:chargino}), above a fairly
smooth SUSY background and a small SM background. Their sum yields the
black points, with the statistical errors also shown. The SUSY
background has a small peak under the signal peak. This is due to a
large number of events with misidentification of $Z$s from neutralino
decays as $W$s, which can be understood from the neutralino branching
ratio in Table~\ref{tab:BRs}. We see immediately the remnant of an
edge effect at $m_{qW}\sim 500$~GeV, as expected from the true
invariant mass distribution shown in Fig.~\ref{fig:imtruth}. However, the
lower edge of the trapezoidal distribution in Fig.~\ref{fig:imtruth}
at $m_{qW}\sim 300$~GeV is not visible, largely as a result of the
hard cuts used on jet momenta. For comparison, we also show in
Fig.~\ref{fig:m_alpha}b the corresponding distribution for the
$qW^\pm$ mass distribution if the cut on the separation scale is {\it
not} imposed: the edge structure is less significant, though still
clearly present, with the SUSY background being larger relative to the
signal.

The $\ell q$ invariant mass distribution resulting from $W^\pm \to
\ell^\pm \nu$ decays is shown in Fig.~\ref{fig:m_alpha}c: here we have
required a lepton ($e$ or $\mu$) with $p_T>40$~GeV instead of the
$W$-jet cuts. Due to the smearing from the escaping neutrino, there is
no interesting edge structure in the signal, and the lack of any
efficient cuts other than the $p_T$ of the lepton means that the
background dominates. While there may be some information to be gained
from the peak positions of the distribution, the leptonic decay is
clearly more difficult to use than the hadronic decay utilising the
\kt\ jet algorithm.

We also show the corresponding invariant mass distribution for $qZ^0$
combinations followed by $Z^0$ decays into hadrons in
Fig.~\ref{fig:m_alpha}d, and for leptonic $Z^0$ decays in
Fig.~\ref{fig:m_alpha}e. For the hadronic decay the small mass
difference between the $W$ and $Z$ means that the signal events (blue)
containing the decay chain (\ref{eq:chi02Z}) are swamped by a large
SUSY background consisting of events with a misidentified $W$, despite
the higher cut values for jet mass and separation scale.

In the case of leptonic $Z$ decays the background can be reduced much
more effectively, by selecting same-flavour, opposite-sign lepton
pairs. As with the jets, this pair is rescaled to the known $Z$
mass. The resulting distribution exhibits the expected signal peak for
$qZ$ combinations (blue), smeared by combinatorial effects from
picking the incorrect jet, over a smaller SUSY background.

Finally, in Fig.~\ref{fig:m_alpha}f we show the invariant $qh$ mass
distribution, where there is, as expected, no signal because of the
zero branching ratio of the $\alpha$ benchmark.

\begin{figure}
\begin{center}
\epsfxsize 14cm
\epsfbox{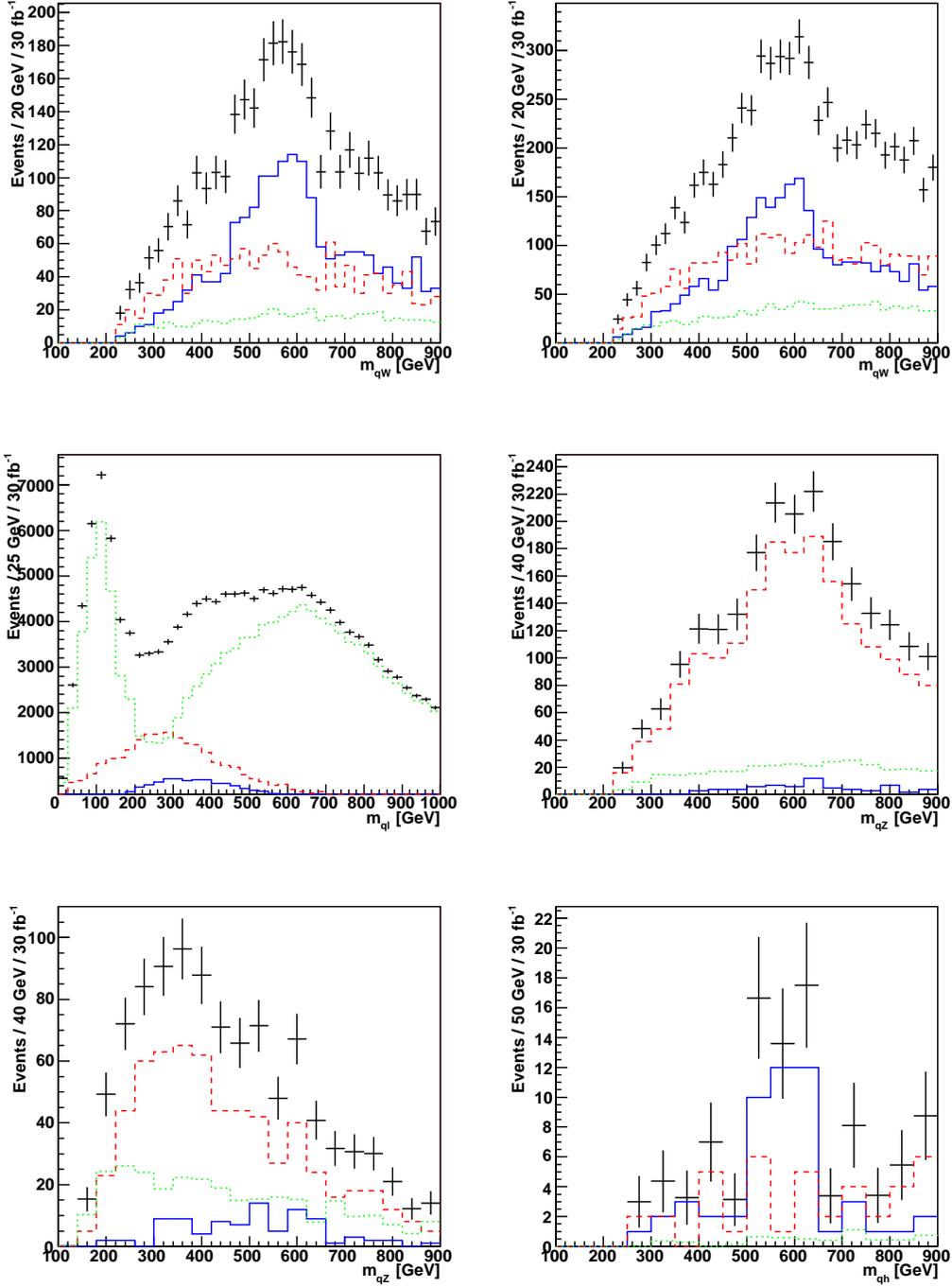}
\caption{Various invariant mass distributions obtained in a simulation
of events for SUSY benchmark scenario $\beta$. See
Fig.~\ref{fig:m_alpha} and text for details.
}
\label{fig:m_beta}
\end{center}
\end{figure}

The corresponding distributions for benchmark $\beta$ are shown in
Fig.~\ref{fig:m_beta}, again for a simulation of 30~fb$^{-1}$. We see
in panel (a) the expected distinctive edge structure (blue, solid)
rising above the SUSY background: the SM background is again small. As
in the case of benchmark $\alpha$, we see in panel (b) that the
signal-to-background ratio is worse if the cut on the separation scale
is {\it not} imposed. We also observe that the SUSY background has
{\it no} visible peak in the signal peak region, due to the small
branching ratio of the neutralino to $Z$ for the $\beta$
benchmark. Panel (c) of Fig.~\ref{fig:m_beta} shows that, as in the
case of benchmark $\alpha$, it would be very difficult to extract
information from a leptonic $W$-decay signal. Nor, according to panels
(d) and (e), does it seem possible in the case of benchmark $\beta$ to
extract a $Z$-decay signal, at least at the considered integrated
luminosity. This might have been anticipated because of the much smaller
$\tilde\chi_2^0\rightarrow\tilde\chi_1^0 Z$ branching ratio in this
case compared to the $\alpha$ benchmark.

For the $qh$ distribution in panel (f) the situation is far better. We
see a clear edge in the distribution in the expected region of
$m_{qh}\sim 650$~GeV. The statistics are low, and are naturally
dependent on the $b$-tagging efficiency. In addition, the observation
of the edge relies on the use of the sub-jet separation scale
cut. Given the good signal-to-background ratio apparent after this
cut, we expect that a lower $b$-tagging efficiency could be
compensated by higher statistics.

\begin{figure}
\begin{center}
\epsfxsize 14cm
\epsfbox{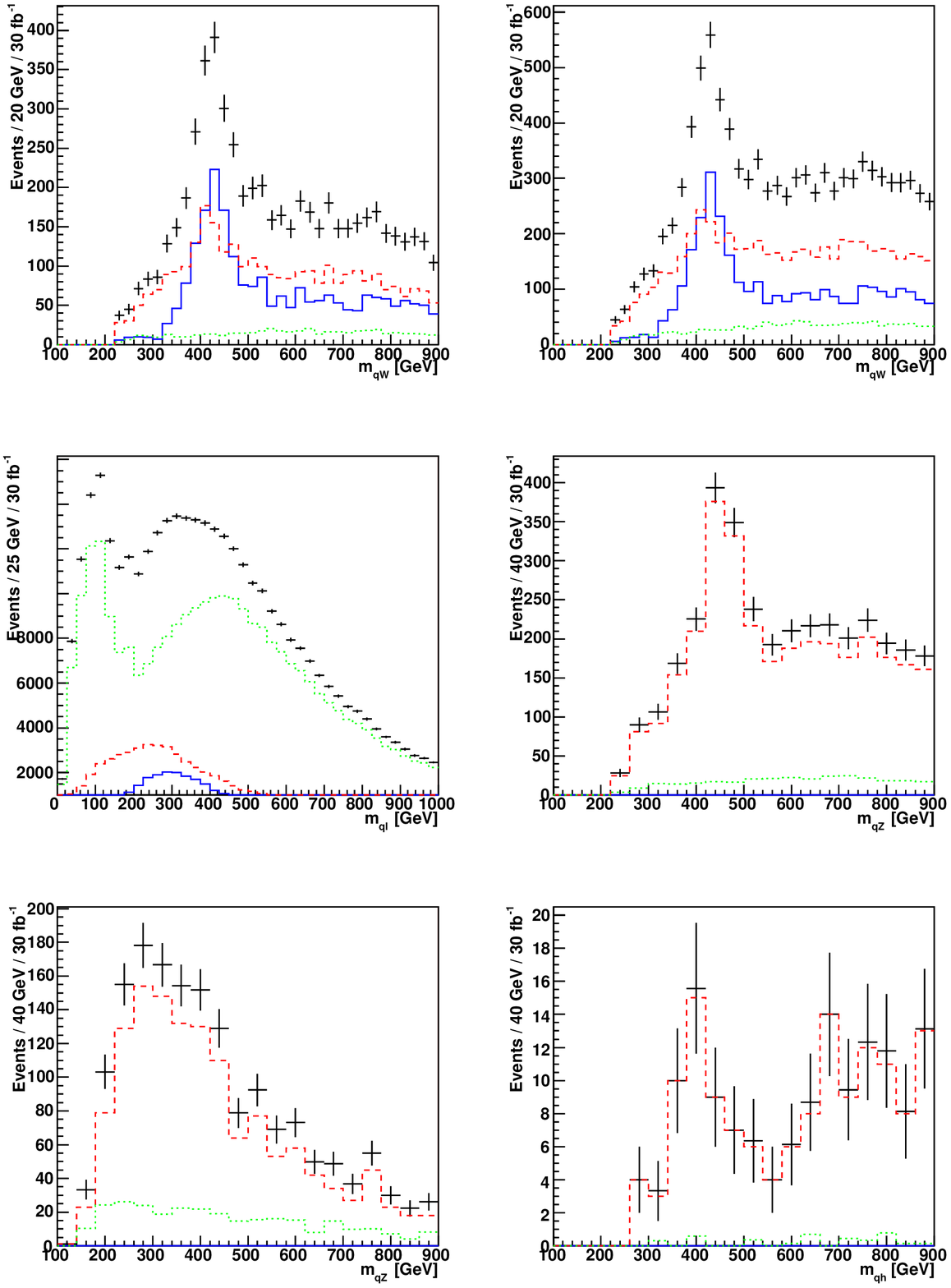}
\caption{Various invariant mass distributions obtained in a simulation
of events for SUSY benchmark scenario $\gamma$. See
Fig.~\ref{fig:m_alpha} and text for details.
}
\label{fig:m_gamma}
\end{center}
\end{figure}

In the case of benchmark $\gamma$, shown in Fig.~\ref{fig:m_gamma},
also for a simulation of 30~fb$^{-1}$, we expect only one observable
distribution. We see again the familiar features of a strong hadronic
$W$-decay signal with the cut on the separation scale in panel (a) and
a weaker signal-to-background ratio without this cut in panel
(b). Again there is a peak in the SUSY background under the signal
peak. For $\gamma$ the on-shell decay
$\tilde\chi_2^0\rightarrow\tilde\chi_1^0 Z$ is not allowed
kinematically, but proceeds off-shell and the decay products are again
misidentified as $W$s. As expected, there are no detectable signals in
leptonic $W$ decay, in hadronic or leptonic $Z$ decay, or in hadronic
$h$ decay.

\begin{figure}
\begin{center}
\epsfxsize 14cm
\epsfbox{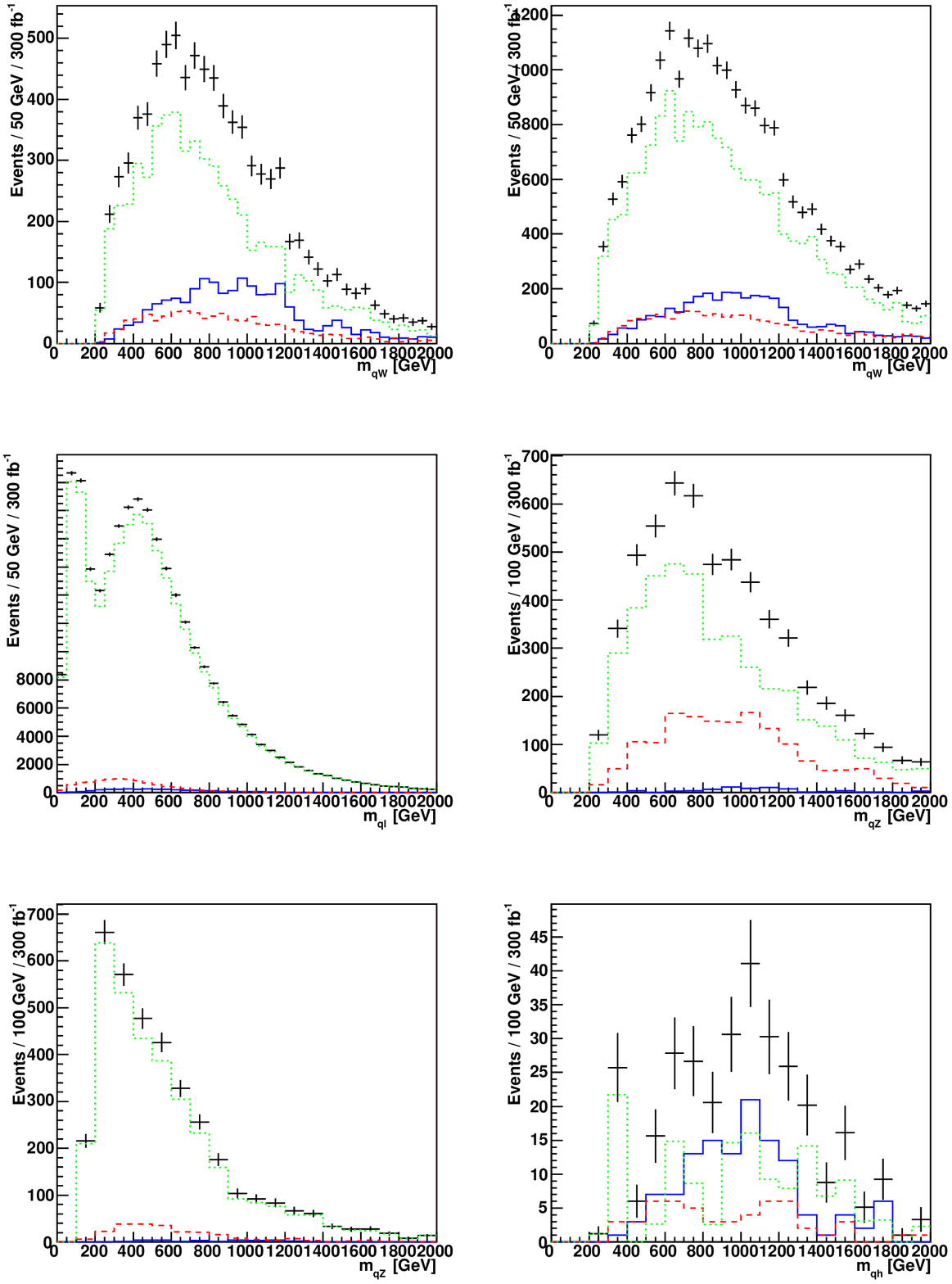}
\caption{Various invariant mass distributions obtained in a simulation
of events for SUSY benchmark scenario $\delta$. See
Fig.~\ref{fig:m_alpha} and text for details.
}
\label{fig:m_delta}
\end{center}
\end{figure}

We turn finally to the case of benchmark $\delta$, shown in
Fig.~\ref{fig:m_delta}. We recall that in this case the simulation
corresponds to an integrated luminosity of 300~fb$^{-1}$, in view of
the higher masses of the sparticles and hence the lower cross
sections. We see in panel (a) that the hadronic $W$-decay signal is
less clear in this case, and that for the first time the SM background
dominates over that due to SUSY. There are virtually no signal events
in panels (c, d) and (e), corresponding to leptonic $W$ decays,
hadronic and leptonic $Z$ decays, respectively. However, there is a
possible $h$ signal in panel (f). The limited amount of generated
events for the SM backgrounds results in relatively large weights for
the background at this integrated luminosity, obfuscating the edge
structure. We again emphasise the necessity of the sub-jet scale cut
and the dependence on the $b$-tagging efficiency assumed.

\subsection{Sideband Subtraction}

\begin{figure}
\begin{center}
\epsfxsize 15cm
\epsfbox{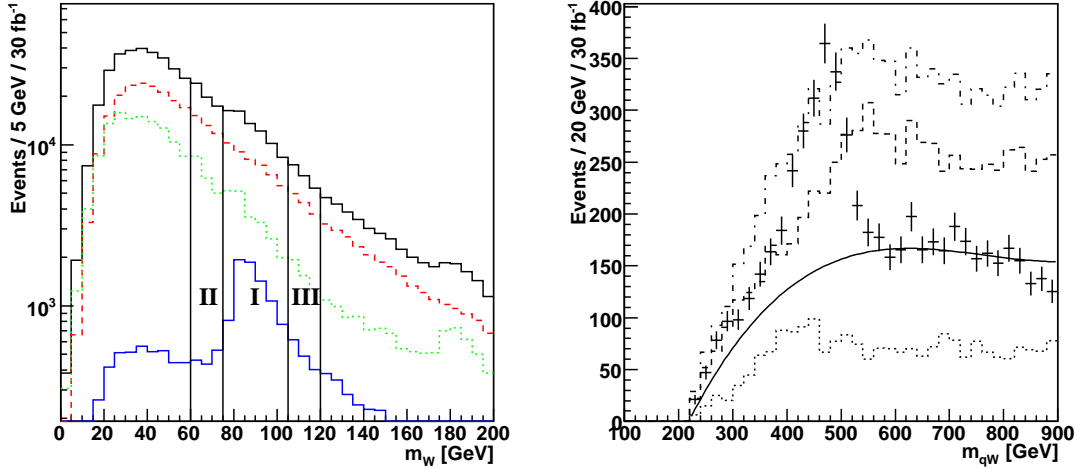}
\caption{Left: Jet mass distribution for $W$ candidates at the $\alpha$
benchmark point following a cut on missing energy:
$\not\!\!E_T>300$~GeV. The SUSY signal (blue, solid), SUSY background
(red, dashed) and SM background (green, dotted) contributions are also
shown separately. Right: Invariant mass distribution of $qW$
combinations for the $\alpha$ benchmark point in the signal region
(region I, points with error bars), in the sideband regions (region II
dashed, region III dotted) and for the sum of sideband events (dashed
dotted). Also shown is the fit to the sideband distribution (solid
line), rescaled to the signal distribution.}
\label{fig:sideband}
\end{center}
\end{figure}

In order to be able to measure the positions of the expected edges of
the invariant mass distributions for signal events, and hence
constrain the sparticle masses, we would like to subtract the SM and
SUSY backgrounds, without model-dependent assumptions on their
shape. We do this by performing a sideband subtraction, where we
imitate the background that does not feature a correctly identified
boson by collecting a sample of events from the generated ``data'',
that features boson candidates with masses away from the resonance
peak of the boson mass in question. Using events in two bands
(region~II and III) on either side of the signal isolation interval
(region~I) for the jet mass distribution, each with half the width of
the signal band, we recalibrate the boson mass to the nominal peak
value as described above, and perform most other cuts as for the
signal. The exception is the sub-jet separation scale cut, which is
highly correlated with the jet mass cut, and is thus ignored for the
sideband sample. We show the jet mass distribution and the signal and
sideband regions for $W$ candidates at the $\alpha$ benchmark point in
Fig.~\ref{fig:sideband} (left). Only the missing-energy cut given in
Section~\ref{sec:cuts} has been applied to the events.

\begin{figure}
\begin{center}
\epsfxsize 15cm
\epsfbox{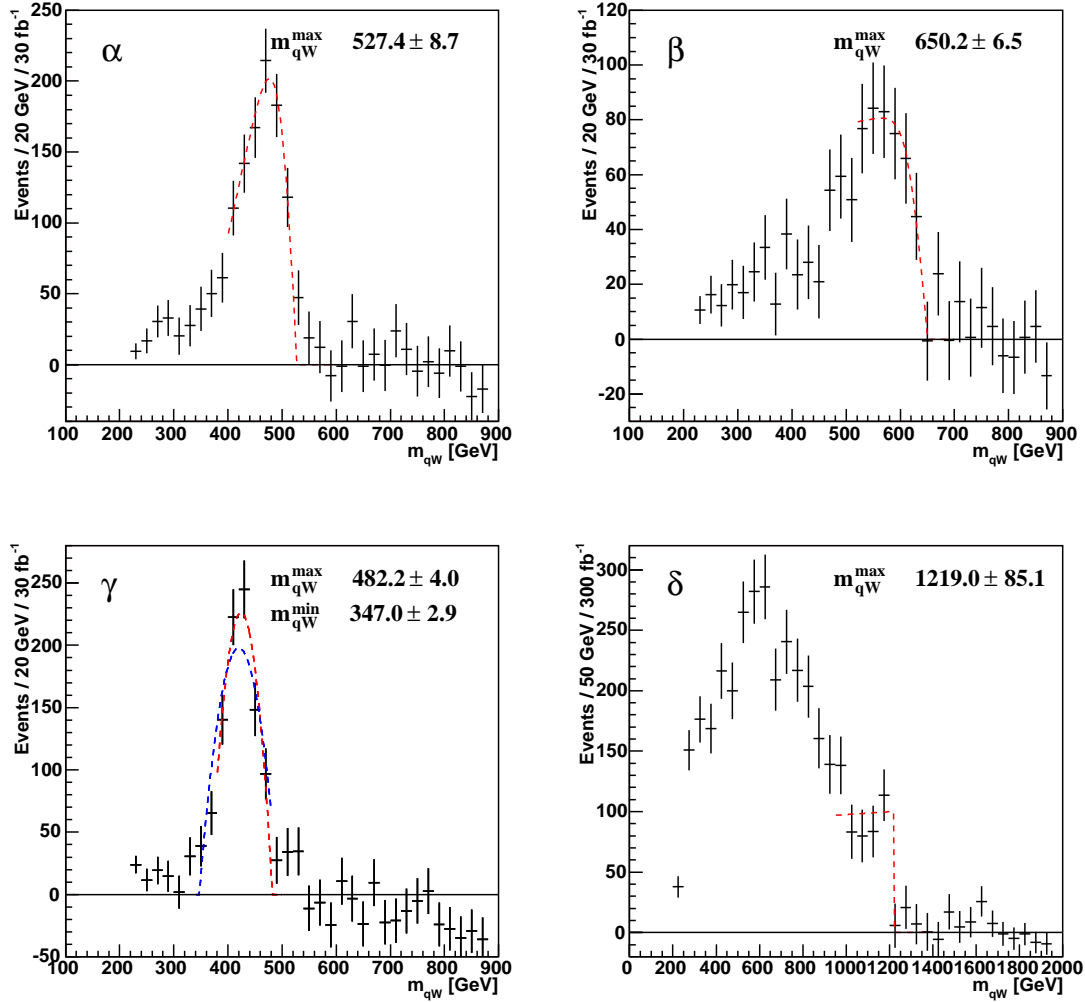}
\caption{Invariant mass distribution of $qW$ combinations at the four
benchmark points, $\alpha$ (top left), $\beta$ (top right), $\gamma$
(bottom left) and $\delta$ (bottom right), after sideband
subtraction. Also shown are fits to the upper edges of the
distributions, and for $\gamma$ also the clear lower edge. }
\label{fig:m}
\end{center}
\end{figure}

The two resulting distributions are added and fitted with a
third-degree polynomial, giving the shape of the background. This is
shown, again using the $\alpha$ benchmark point as an example, in
Fig.~\ref{fig:sideband} (right). The background is rescaled to the
full distribution from the signal band, shown with error bars in
Figs.~\ref{fig:m_alpha}-\ref{fig:m_delta}, using bins at higher
invariant masses than the observed edges. The rescaled background is
then subtracted from the full distributions, the results of which are
shown for the $qW$ invariant mass distributions in
Fig.~\ref{fig:m}. While this procedure primarily models background
from fake $W$s, we find that it does a good job of describing all the
background near the upper edges, perhaps with the exception of the
$\delta$ benchmark point which is dominated by SM events with real
$W$s.  For the other three benchmarks, with much less SM background,
the jet rescaling to the $W$ mass for the sideband samples gives a
distribution that is similar enough to model well also this background.

We have also considered estimates for the combinatorial background
that result from combining correctly identified bosons with the wrong
jet, which can be seen as tails in the blue signal distributions in
Figs.~\ref{fig:m_alpha}-\ref{fig:m_delta}. By combining accepted boson
and squark decay jet candidates randomly from all events we collect a
large mixed sample that should be representative of this part of the
background. However, we find that combining these two descriptions of
the background is difficult due to the limited statistics, and joint
fits to the events in the high invariant-mass region tend to favour
heavily the sideband sample. For the $\delta$ benchmark point the
sideband and mixed samples perform similarly in describing the
background in the vicinity of the edge, but the sideband is slightly
better at the high invariant masses used to set the scale of the
background distribution. As a consequence, the mixed sample is not
included in the fits shown in this Section, but investigations into
combining different descriptions of the background is certainly worthy
of further effort, in particular when data is available from the
experiments.
 
\begin{figure}
\begin{center}
\epsfxsize 15cm
\epsfbox{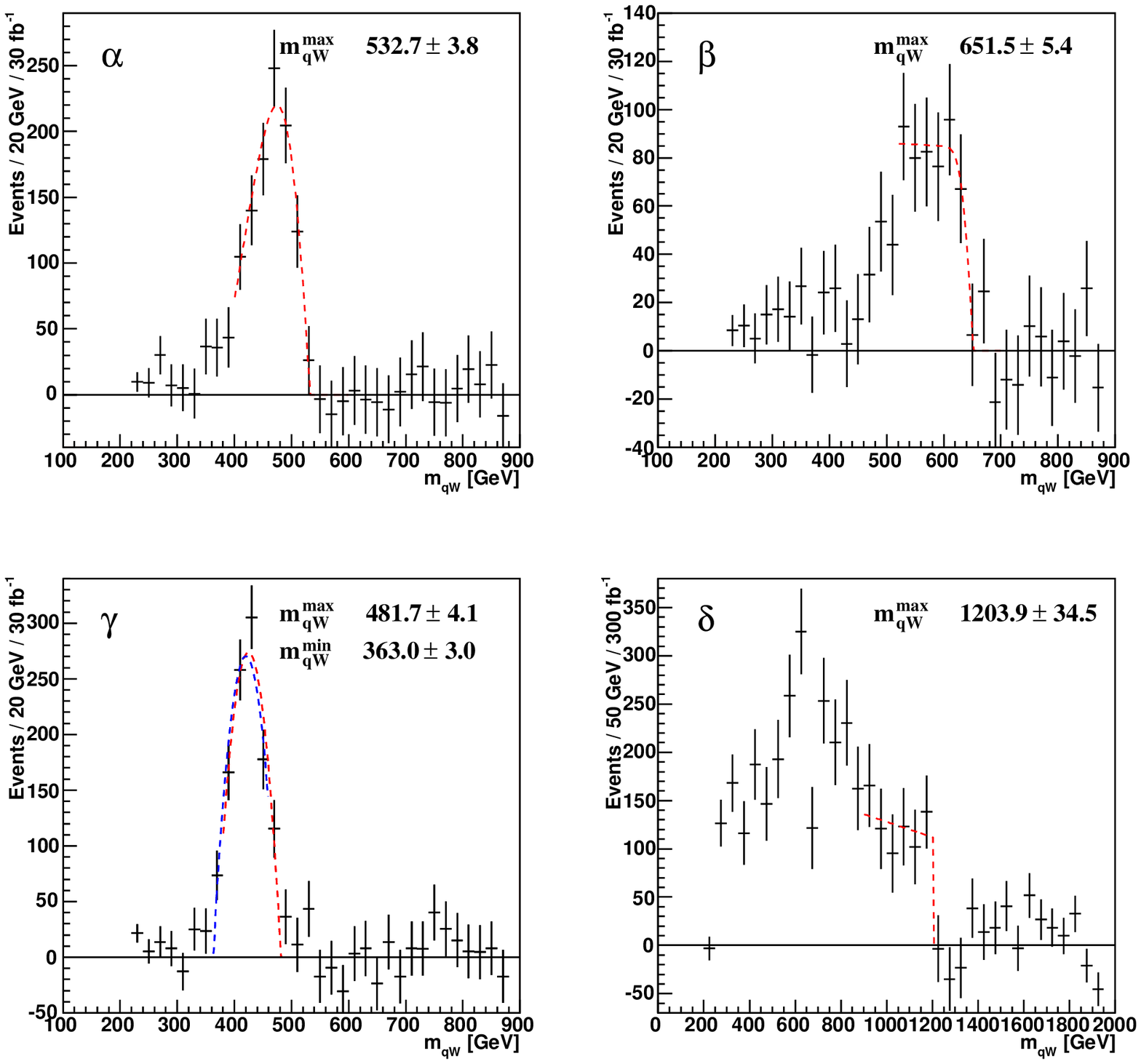}
\caption{As Fig.~\ref{fig:m}, but omitting the cut on the sub-jet
separation scale.}
\label{fig:m_nosep}
\end{center}
\end{figure}

After the sideband subtraction, the upper endpoints of the $qW$
distributions are clearly visible for all four benchmarks in
Fig.~\ref{fig:m}. To estimate their positions, we have performed fits
with a linear form for the signal, emulating the distributions of
Fig.~\ref{fig:imtruth}, with a free parameter for the cut-off at the
endpoint. These distributions are further smeared by a Gaussian to
model the limited jet energy resolution, using a smearing width
determined by the fit. The resulting values for the endpoints
$m_{qW}^{\rm max}$ can be found in Table~\ref{tab:qWmax}. We also show
in Fig.~\ref{fig:m_nosep} the $qW$ invariant mass distributions obtained
when we omit the cut on separation scale for the $W$ candidate, and
the corresponding fit values are also reported in
Table~\ref{tab:qWmax}.

\begin{table}\begin{center}
\begin{tabular}{lcccc} \hline
Fit / Benchmark & $\alpha$ & $\beta$ & $\gamma$ & $\delta$ \\ \hline
Scale cut & $527.4\pm 8.7$ & $650.2\pm 6.5$ & $482.2\pm 4.0$ & $1219.0\pm 85.1$ \\
No scale cut & $532.7\pm 3.8$ & $651.5\pm 5.4$ & $481.7\pm 4.1$ & $1203.9\pm 34.5$ \\
Nominal      & $519.9$ & $653.8$ & $468.6$ & $1272.1$ \\ \hline
\end{tabular}
\caption{Fitted endpoint values $m_{qW}^{\rm max}$ of $qW$ distributions,
measured in GeV, compared with the nominal values for the
corresponding benchmarks.}
\label{tab:qWmax}
\end{center}
\end{table}
 
 
 
 
 
We see that the endpoint estimates have statistical errors which are
in the ${\cal O}(1)$\% region for benchmarks $\alpha-\gamma$ and
slightly below ${\cal O}(10)$\% for $\delta$. The statistical errors
are in general smaller without the cut on the sub-jet separation
scale, particularly in the case of benchmark $\delta$, due to somewhat
larger statistics.

In the case of the $\alpha$ benchmark point, the fitted values are
both fairly close to the nominal value, but with indications of larger
systematic errors when not using the scale cut. Both fits overestimate
the endpoint, which could be expected as the reconstructed jets from
squark decays have not been calibrated for the algorithm's tendency to
overestimate the jet. For $\beta$ both fits are very close to the
nominal value, while for $\gamma$ the fits again overestimate the edge
position. From the three benchmarks considered here this systematic
error seems to increase for lower values of the endpoint, barring
other more important hidden systematics. In the case of the $\delta$
benchmark both fits underestimate the endpoint, but here the
statistical errors are large. Considered as a whole, the results of
Table~\ref{tab:qWmax} indicate that the systematical errors on the
endpoints in our fitting procedure are comparable to the statistical
errors.

For most of the distributions it is impossible to estimate the lower
endpoint, since any structure at low invariant masses is washed away
by the hard kinematical cuts on jet energies used to isolate the
signal. The exception is the relatively high lying lower endpoint of
the $qW$ distribution for the $\gamma$ benchmark, where a similar
fitting technique to the one used for the upper endpoint yields a
lower endpoint estimate of $m_{qW}^{\rm min}=363.0\pm 3.0$~GeV before
and $m_{qW}^{\rm min}=347.0\pm 2.9$~GeV after the sub-jet cut is
applied, compared to the nominal value of $369.0$~GeV. The difference
in endpoint estimate originates from a worse fit of the sideband
distribution at high invariant masses when the sub-jet cut is used.

We have in addition estimated both the lower and upper edge of the
$qZ$ distribution for the $\alpha$ benchmark point, where we
reconstruct the $Z$ from a lepton pair. After a similar sideband
subtraction we estimate $m_{qZ}^{\rm max}=523.7\pm 10.6$~GeV and
$m_{qZ}^{\rm min}=324.5\pm 9.2$~GeV, to be compared to the nominal
values of $505.6$~GeV and $358.5$~GeV, respectively.

For the $qh$ distributions, fitting is difficult with the low
statistics available both at the $\beta$ and $\delta$ benchmarks. From
sideband subtracted distributions we make estimates by visual
inspection, giving values of $m_{qh}^{\rm max}=540\pm 40$~GeV for
$\beta$ and $m_{qh}^{\rm max}=1450\pm 100$~GeV for $\delta$, compared
to the nominal values of $628.3$~GeV and $1265.7$~GeV, respectively. As
has been noted earlier, the feasibility of measuring these edges
depends on the $b$-tagging achievable, and on the use of the sub-jet
separation scale cut.

\subsection{Mass Spectra}
\label{sec:mass}

\begin{figure}
\begin{center}
\epsfxsize 15cm
\epsfbox{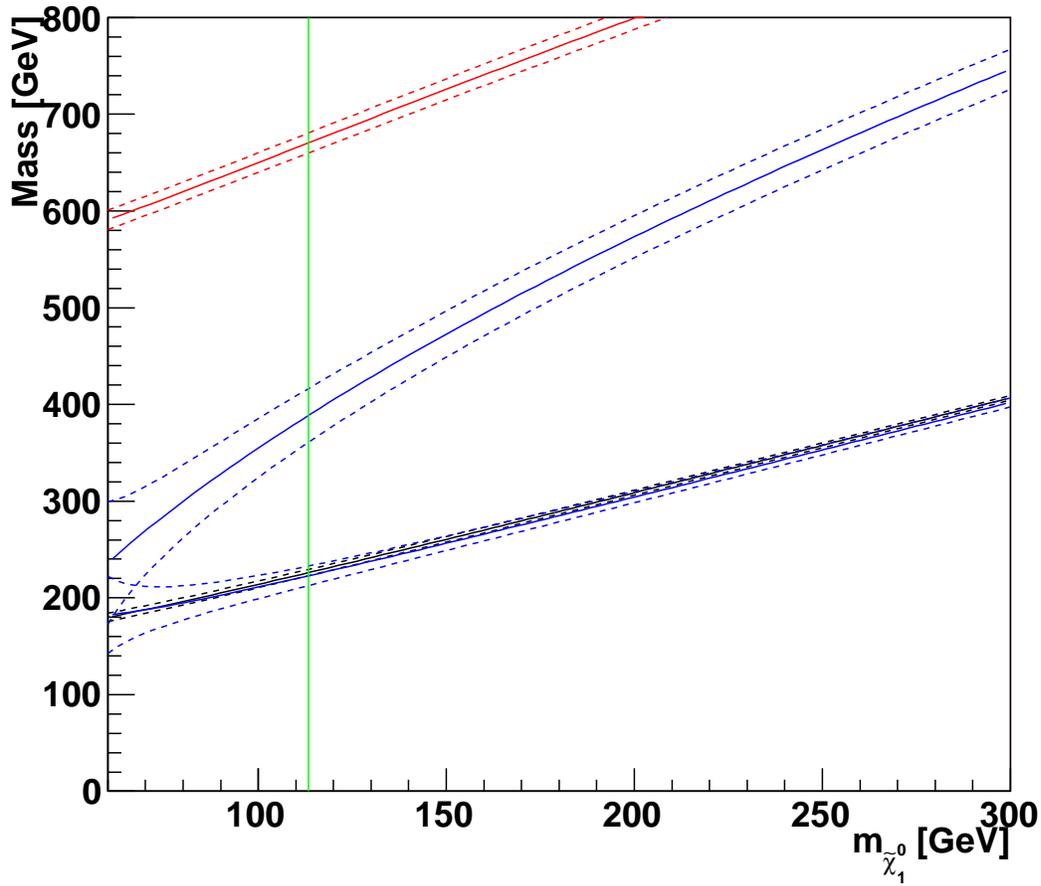}
\caption{The squark (red), $\tilde\chi_1^\pm$ (blue) and
$\tilde\chi_2^0$ (black) masses as a function of the
$\tilde\chi_1^0$ mass for the $\alpha$ benchmark point. The dashed
lines show the $1\sigma$ statistical error bands. The vertical green
line indicates the nominal $\tilde\chi_1^0$ mass.}
\label{fig:masses_alpha}
\end{center}
\end{figure}

The positions of the edges of the invariant mass distributions may be
used to extract information on the spectrum of the SUSY particles
involved. Whilst the four equations giving the edges of the $qW$ and
$qZ/qh$ distributions may in principle be solved to obtain the four
masses involved, extracting accurate values of the absolute masses -
as opposed to mass differences - will be difficult in practise because
of the degeneracies between the masses of the bosons and of the
gauginos, respectively, and thus in the upper edges, as shown in
Fig.~\ref{fig:imtruth}. However, for the four benchmarks considered
here this is an academic problem, as neither of the benchmarks have four
measurable edges. Given both the lower and upper edge of a single
distribution the squark mass can be eliminated from
Eq.~\ref{eq:endqW}, and one can solve for the mass of the chargino or
the next-to-lightest neutralino in terms of the mass of the lightest
neutralino. This can in turn be used to arrive at the squark mass. We
show in Fig.~\ref{fig:masses_alpha} the squark and $\tilde\chi_2^0$
masses determined using the upper and lower $qZ$ edge measured for the
$\alpha$ benchmark point, as functions of the undetermined
$\tilde\chi_1^0$ mass. We also show the $1\sigma$ error bands on the
masses resulting from the statistical uncertainty of the edge
measurements, assuming that the errors on the two edges are
independent.

With the squark mass known in terms of the $\tilde\chi_1^0$ mass, the
upper edge of the $qW$ distribution can also be used to give the
$\tilde\chi_1^\pm$ mass. Because of the quadratic nature of
Eq.~\ref{eq:endqW} two solutions result, none of which can be rejected
out of hand~\footnote{There are also in principle two solutions for
the $\tilde\chi_2^0$ mass, but one can be rejected as unphysical since
it is always less than the $\tilde\chi_1^0$ mass.}. We show both
solutions in Fig.~\ref{fig:masses_alpha}. For values of the
$\tilde\chi_1^0$ mass below $\sim 50$~GeV there are no solutions.

\begin{figure}
\begin{center}
\epsfxsize 15cm
\epsfbox{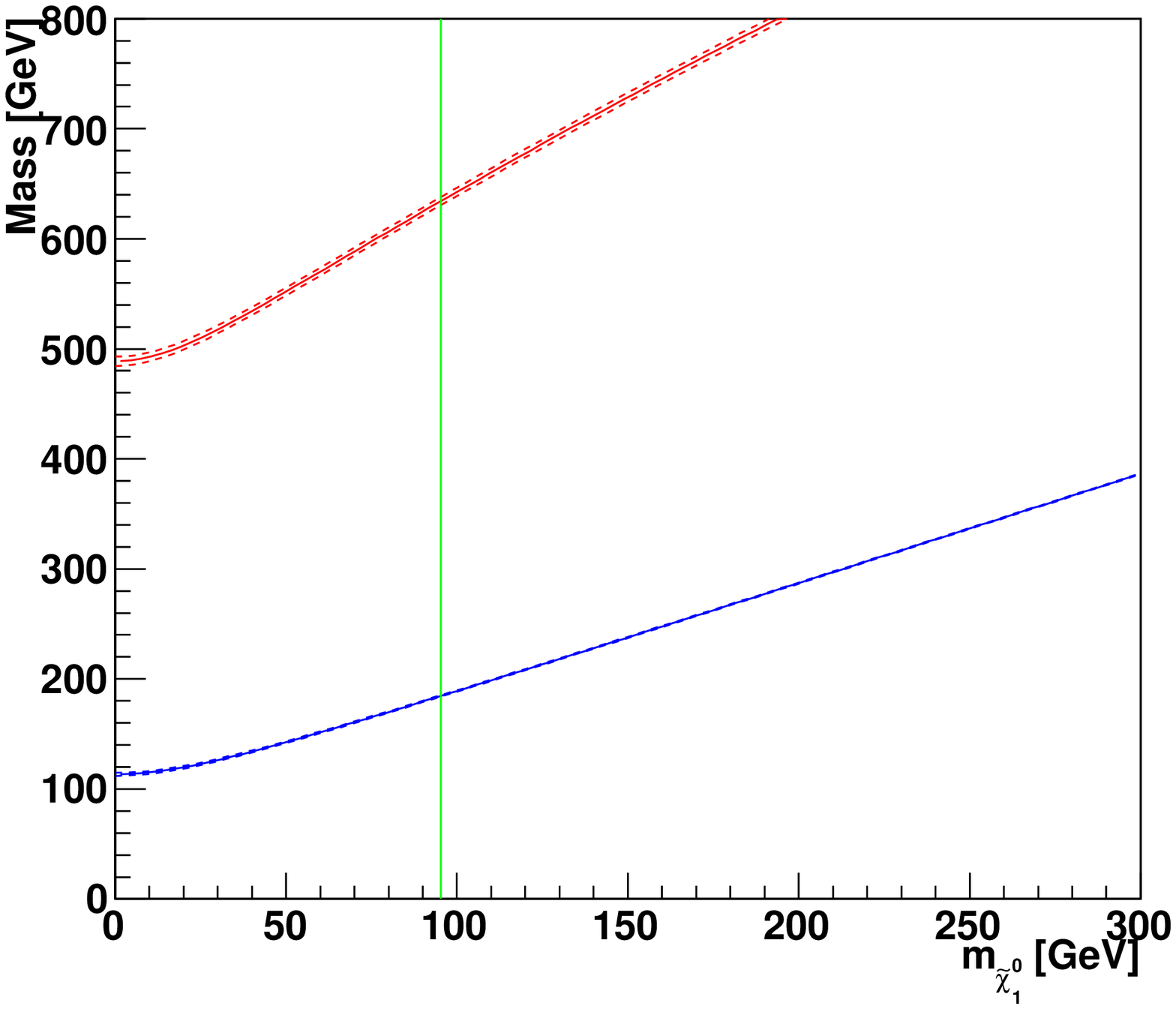}
\caption{The squark (red) and $\tilde\chi_1^\pm$ (blue) and
masses as a function of the $\tilde\chi_1^0$ mass for the $\gamma$
benchmark point. The dashed lines show the $1\sigma$ statistical error
bands. The vertical green line indicates the nominal $\tilde\chi_1^0$
mass.}
\label{fig:masses_gamma}
\end{center}
\end{figure}

For the $\gamma$ benchmark point the same procedure can be followed
for the two endpoints of the $qW$ distribution, and we show the
resulting squark and chargino masses as functions of the lightest
neutralino mass in Fig.~\ref{fig:masses_gamma}. For the remaining two
benchmarks, extracting masses is more difficult since only one edge is
well measured. However, the LHC experiments have the potential for
measuring other SUSY mass-dependent quantities, such as the effective
mass~\cite{Hinchliffe:1996iu,Tovey:2000wk}, whose determination would
give complementary relations between the involved masses.

Given an accurate measurement of the $\tilde\chi_1^0$ mass, for
example at a future linear collider, the heavier neutralino, chargino
and squark masses could then be found with statistical errors in the
range of $1-5$\%, using LHC data, even if the particles themselves are
too heavy to be produced at the linear collider.

\section{Results and Conclusions}
\label{sec:results}

We have shown in this paper, we believe for the first time, that it is
possible to extract a SUSY signal solely from the hadronic decays of
$W^\pm$ bosons produced in cascade decays involving charginos, and
that the measurement of this signal may provide useful information
about the sparticle mass spectrum. In each of the benchmarks studied,
the upper edge of the $qW$ mass distribution in the decay chain
$\tilde q_L\rightarrow\tilde\chi_1^\pm q \rightarrow\tilde\chi_1^0
W^\pm q$ can be measured, with a statistical error of ${\cal O}(1)$\%
for the relatively light sparticle masses of benchmarks
$\alpha-\gamma$ and ${\cal O}(10)$\% for the heavier $\delta$, with
squark and gluino masses above $1.5$~TeV. Systematic errors in the
background and fitting are estimated to be of the same order as the
statistical errors. In the case of benchmark $\gamma$, we are also
able to extract the position of the lower edge of the $qW$ mass
distribution, and in the case of point $\alpha$ we extract both the
upper and the lower edges of the $qZ$ mass distribution, using the
leptonic decays of $Z$ bosons. There are also clear indications in
benchmarks $\beta$ and $\delta$ for the possibility of observing the
upper edge of the $qh$ mass distribution using $h \to {\bar b} b$
decays with collimated $b$-jets, but conclusions on such a signal
require further analysis and better understanding of the $b$-tagging
efficiency for these jets in a full detector simulation.

The hadronic signals were extracted using the \kt\ algorithm for jet
reconstruction, including the single-jet mass and the proposal
of~\cite{Butterworth:2002tt} to improve the identification of hadronic
decays of heavy bosons via a cut on the sub-jet separation scale. The
\kt\ algorithm is shown to be well suited for reconstructing edge
features in the considered invariant mass distributions. The sub-jet
cut procedure improves quantitatively the signal-to-background ratio,
and while the loss of statistics increases the statistical errors,
fits to the edges of the $qW$ distribution show improved results with
respect to the nominal values when using the sub-jet cut. For the $qh$
distributions with their small signal-to-background ratio, the sub-jet
cut is crucial in reducing the background to a manageable level.

A detailed exploration of the capabilities of the LHC experiments to
measure and use all the information that could be gained from these
benchmark points, and hence a more detailed display of the value added
to the global fits by the measurements described here, lies beyond the
scope of this study. However, we do note that this technique provides
novel information on chargino and neutralino spectroscopy, e.g., the
difference between the $\tilde
\chi_1^0$ and $\tilde\chi_1^\pm$ mases, and in at least one case the
$\tilde\chi_2^0 - \tilde\chi_1^\pm$ mass difference. These pieces of
information are useful for potentially constraining SUSY models and
perhaps foreseeing the locations of interesting thresholds in $e^+
e^-$ annihilation.

As pointed out in~\cite{Butterworth:2002tt}, similar techniques for
analysing hadronic final states arising from the decays of heavy
particles should be useful in other situations. Examples include the
analysis of top physics, the search for $R$-violating hadronic decays
of sparticles, or in the isolation of a generic SUSY signal from QCD
SM backgrounds. Indeed, some recent studies have highlighted the
potential of single-jet mass cuts in exotic searches~\cite{studies},
and we note that the $y$-scale cut should also be useful in these
cases. We believe this to be an area meriting much further
experimental and phenomenological study.

\section*{Acknowledgements}
The authors thank Jeppe Anderson for useful discussions. ARR
acknowledges support from the European Community through a Marie Curie
Fellowship for Early Stage Researchers Training, and from the
Norwegian Research Council.

\end{document}